\documentclass[a4paper,11pt]{article}

\usepackage[margin=1.14in]{geometry} 
\usepackage{tabularx}
\usepackage{multirow}
\usepackage{hyperref}

\usepackage{arydshln}
\usepackage{amsmath}
\usepackage{amsthm}
\usepackage{amssymb}
\usepackage{graphicx}
\usepackage{pdfpages}
\usepackage{float}
\makeatletter
\makeatletter \renewcommand{\@dotsep}{10000} \makeatother
\usepackage{wrapfig}
\usepackage[utf8]{inputenc}
\usepackage{lmodern}
\usepackage{hyperref}
\hypersetup{
	unicode,
	colorlinks,
	breaklinks,
	urlcolor=cyan, 
    linkcolor=black, 
	pdfauthor={Author One, Author Two, Author Three},
	pdfproducer={LaTeX},
	pdfcreator={pdflatex}
}
\usepackage[T1]{fontenc} 
\usepackage{amsmath}

\newcommand{\be}{\begin{eqnarray}}
\newcommand{\ee}{\end{eqnarray}}
\def\be{\begin{equation}}
\def\ee{\end{equation}}
\def\bea{\begin{eqnarray}}
\def\eea{\end{eqnarray}}

\newcommand{\gsim}{\;\raisebox{-0.9ex}{$\textstyle\stackrel{\textstyle >}{\sim}$}\;}
\newcommand{\lsim}{\;\raisebox{-0.9ex}{$\textstyle\stackrel{\textstyle<}{\sim}$}\;}
\def\lsim{\raise0.3ex\hbox{$\;<$\kern-0.75em\raise-1.1ex\hbox{$\sim\;$}}}
\def\gsim{\raise0.3ex\hbox{$\;>$\kern-0.75em\raise-1.1ex\hbox{$\sim\;$}}}

\usepackage{graphics}
\def\met{\slashed E_T}
\usepackage{epsfig}
\usepackage{slashed}
\usepackage[utf8]{inputenc}
\usepackage{multirow}
\usepackage{pstricks}
\usepackage{dcolumn}

\usepackage[sort&compress,numbers,square]{natbib}

\theoremstyle{plain}

\theoremstyle{definition}

\usepackage{graphicx, color}

\title{Testing CP properties of the Higgs boson coupling to $\tau$ leptons with heterogeneous graphs}
\vspace{8mm}
\author{\Large{W. Esmail$^{a}$,  A. Hammad$^{b}$, M. Nojiri$^{b,c,d}$ and Christiane Scherb$^{e,f}$}}

\date{
{}
$^a$ Institut f\"ur Kernphysik, Universit\"at M\"unster, 
Wilhelm-Klemm-Str.\\  9, 48149 M\"unster, Germany.\\
$^b$ Theory Center, IPNS, KEK,  1-1 Oho, Tsukuba, Ibaraki 305-0801, Japan.\\
$^c$ The Graduate University of Advanced Studies (Sokendai),\\ 1-1 Oho, Tsukuba, Japan.\\
$^d$ Kavli IPMU (WPI), University of Tokyo, 5-1-5 Kashiwanoha,\\ Kashiwa, Chiba 277-8583, Japan.\\
$^e$ Berkeley Center for Theoretical Physics, Department of Physics,\\
University of California, Berkeley, CA 94720, USA.\\
$^f$ Theoretical Physics Group, Lawrence Berkeley National Laboratory,\\ Berkeley, CA 94720, USA. }

\begin{document}
	\maketitle
	\vspace{4mm}
	\begin{abstract}
 \normalsize{
We explore the feasibility of measuring the CP properties of the Higgs boson coupling to $\tau$ leptons at the High Luminosity Large Hadron Collider (HL-LHC). Employing detailed Monte Carlo simulations, we analyze the reconstruction of the angle between $\tau$ lepton planes at the detector level, accounting for various hadronic $\tau$ decay modes. Considering standard model backgrounds and detector resolution effects, we employ three Deep Learning (DL) networks, Multi-Layer Perceptron (MLP), Graph Convolution Network (GCN), and Graph Transformer Network (GTN) to enhance signal-to-background separation. To incorporate CP-sensitive observables into Graph networks, we construct Heterogeneous graphs capable of integrating nodes and edges with different structures within the same framework. Our analysis demonstrates that GTN exhibits superior efficiency compared to GCN and MLP. Under a simplified detector simulation analysis, MLP can exclude CP mixing angle larger than $20^\circ$ at $68\%$ confidence level (CL), while GCN and GTN can achieve exclusions at $90\%$ CL and $95\%$ CL, respectively with  $\sqrt{s}=14$~TeV and $\mathcal{L}=100\rm { fb}^{-1}$.  Furthermore, the DL networks can achieve a significance of approximately $3\sigma$ in excluding the pure CP-odd state.
 }
\end{abstract}
\newpage
\noindent\rule{\textwidth}{1pt}
\tableofcontents
\noindent\rule{\textwidth}{0.2pt}
\maketitle \flushbottom
\vspace{4mm}
\section{Introduction}
\label{sec:intro}

Following the discovery of a scalar resonance with a mass near 125 GeV \cite{ATLAS:2012yve,CMS:2012qbp}, researchers are now focused on measuring the properties and characteristics to assess whether it aligns with the Higgs boson predicted by the Standard Model (SM) \cite{ATLAS:2016neq,ATLAS:2019nkf}. A key aspect of this investigation is examining its spin and CP transformation properties \cite{DellAquila:1985jin,Gao:2010qx,Choi:2002jk}. 
For example, top-associated production is a promising channel to extract such properties
\cite{Buckley:2015vsa,Barman:2021yfh,Esmail:2024gdc}.
Testing its CP properties is especially significant in light of the observed baryon asymmetry of the universe. The SM CP violation has been observed initially in Kaon decays \cite{Christenson:1964fg}, and established by the measurements of the direct CP violation in K system \cite{KTeV:1999kad},  and  CP violation in neutral B meson decays \cite{Belle:2001zzw,BaBar:2001pki}. 
However, it is insufficient to account for the Bayron asymmetry of the Universe  \cite{Sakharov:1967dj}. 
Therefore, additional sources of CP violation are essential ingredients of BSM to address the origin of the matter in our Universe. 
So far, measurements of the Higgs boson properties, for instance, the interactions with gauge bosons \cite{ATLAS:2016ifi,ATLAS:2017azn,ATLAS:2018hxb,CMS:2016tad,CMS:2019ekd,CMS:2019jdw}, performed by the ATLAS and CMS experiments show no deviations from the SM predictions. Still, the possibility of an extended scalar sector that includes CP violation, and thus, 
the observed scalar resonance being a CP mixing state, has not been ruled out.  

Besides the top-associated Higgs production, studying the $\tau$ spin correlation 
of the $h\tau\tau$ coupling yields valuable information about the CP state of the Higgs boson and has been proposed for both the LHC and lepton colliders \cite{DellAquila:1988bko, Bower:2002zx,Desch:2003mw, Berge:2008wi, Berge:2008dr, Berge:2011ij, Berge:2013jra, Harnik:2013aja, Dolan:2014upa, Berge:2015nua, Askew:2015mda, Hagiwara:2016zqz, Antusch:2020ngh}.  
Here, the primary focus is on reconstructing acoplanarity angle of the decays of the $\tau$ lepton pairs from the angular correlation of the decay products. 
A significant challenge in this kind of study at LHC arises from the presence of neutrinos in $\tau$ lepton decays, which complicates the accurate determination of the $\tau$ momentum vector and, consequently, the angular distribution between the $\tau$ lepton pairs at the LHC. 
However, the one-prong decays of the $\tau$ lepton, $\tau^\pm \rightarrow \pi^\pm  \nu$, $\tau^\pm\to \rho^\pm (\rho^\pm\to\pi^\pm\pi^0 ) \nu_\tau, \tau^\pm\to a_1^\pm (a_1^\pm\to\pi^\pm\pi^0\pi^0 ) \nu_\tau$, are promising. 
For the $\pi^+$ final state,  the CP information can be reconstructed from charged pion momentum and interaction point (IP) information \cite{Berge:2008dr}. 
For the case of $\rho$ and $a_1$ final states, 
the angle between the decay planes of ($\pi^+,n\pi^0$), ($\pi^-,n\pi^0$) retains the spin correlation information of the $\tau$ lepton pair for $\rho$ and $a_1$ mode. 
It is essential to combine as many final states as possible; the branching ratio of $\tau^+\to\pi^+\nu $ final state is only  10.8\% of the total decay width but increases up to  45.6\% if one includes up to $\pi^+ 2\pi^0$ final state. 
Analyses of Higgs boson decays into $\tau$ lepton pairs have been conducted by both ATLAS \cite{ATLAS:2015xst, ATLAS:2016ifi,ATLAS:2022akr} and CMS \cite{ CMS:2014wdm,CMS:2017zyp,CMS:2021sdq,CMS:2022uox}. 
They show agreement with the SM prediction of the Higgs boson but do not rule out CP admixture. ATLAS and CMS will accumulate more than 200 fb$^{-1}$ for each experiment in Run 3 (2024-2025), in the future, the High Luminosity Large Hadron Collider (HL-LHC) aims to collect 3000 fb$^{-1}$, to reveal the nature of the Higgs boson.

The hadronic final states of 
$\tau\bar\tau$ suffers serious background from various SM processes. Cuts to reduce the background have been developed, such as the isolation of the jet, jet mass, change multiplicities, and momentum distribution of pions.  
{Deep Learning Neural Networks (DNNs) can efficiently increase the signal-to-background yield.} 
Recently, DNNs have been widely used in collider analysis for various tasks, see  \cite{Feickert:2021ajf} and references therein. In this paper, we utilize different sets of DNNs to suppress background events and study the CP properties of the Higgs boson at the  HL-LHC. The first DNN we consider is the Multi-Layer Perceptron (MLP), which analyzes high-level kinematic and CP distribution. Although MLP can achieve high classification performance between signal and background events, it does not provide optimal performance for an analysis of the CP properties of the Higgs boson. This is because 
CP information is fully mixed with event kinematics; 
the learned information about Higgs boson CP properties will be diluted and network performance will be degraded. 
The issue of degraded performance in MLPs due to inputs containing mixed information has been highlighted, for example, in \cite{Ban:2023jfo}. 

Alternatively, one can utilize a heterogeneous graph to analyse the 
$h\rightarrow \tau\bar\tau$. A heterogeneous graph consists of nodes and edges with different types of information, which allows the separate encoding of CP and kinematical information.  
We construct a graph
of the nodes of the final state pions and the reconstructed $\tau$ pair with selective connections. Pion nodes are fully connected  to recover the kinematic information, while tau nodes are connected with edges weighted by the value of the reconstructed angular distribution between them. This approach allows CP and kinematic information to be separately encoded within a single graph. 

To analyze these constructed graphs, we utilize two Graph Neural Networks (GNNs): Graph Convolutional Network (GCN) and Graph Transformer Network (GTN). One key advantage of the GTN is its ability to dynamically capture the complex information within the graph, as GTNs leverage the attention mechanism to weigh the importance of different nodes irrespective of their position, while GCNs rely on localized neighborhood aggregation. 
Thanks to the attention mechanism, GTNs can dynamically focus on the most relevant parts of the heterogeneous graph. 
This flexibility, together with the enhanced capability to model intricate patterns, make GTNs particularly powerful for heterogeneous graph classification tasks.
We find that GTN can enhance the current sensitivity on CP nature of ATLAS and CMS analysis.  
 
This paper is organized as follows: In Section \ref{sec:2}, we describe the parameterization of the effective Lagrangian relevant to this study and the construction of the CP observable in different hadronically decaying channels of the $\tau$ lepton. Section \ref{sec:3} details the analysis methodology and the Monte Carlo tools used for event simulation. In Section \ref{sec:4}, we discuss the three deep learning methods,  MLP, GCN and  GTN. The results of this study are presented in Section \ref{sec:5} and our conclusion  is drawn in Section \ref{sec:6}.

\section{Effective Lagrangian and CP observable}
\label{sec:2}
The general form of the effective Yukawa interaction between the Higgs boson $h$ and $\tau$ leptons can be written as
\begin{equation}
    \mathcal{L}_{H\tau\tau} = -\frac{m_\tau}{\upsilon}\kappa_\tau\Bar{\tau}\left(\cos\theta_\tau+i\gamma_5\sin\theta_\tau \right)\tau h \,.
    \label{eq:1}
\end{equation}
Here, $\kappa_\tau$ is the reduced Yukawa coupling strength, $v=246$ the vacuum expectation value (vev) of $h$ and $\theta_\tau$ is the CP mixing angle, with $\theta_\tau$ ranging from $-90^\circ$ to $90^\circ$. This angle parameterized the relative contributions of the CP-even and CP-odd components to the $h\tau\tau$ coupling. Specifically, $\theta_\tau = 0^\circ$ represents a purely CP-even state, while $\theta_\tau = 90^\circ$ represents a purely CP-odd state. Values of $\theta_\tau$ between these extremes indicate an admixture of both components, suggesting a CP-violation in the Higgs sector.
The CP-mixing angle $\theta_\tau$ affects the correlations between the transverse spin components of $\tau$-leptons in $h \to \tau \tau$ decays. These correlations, in turn, influence the directions of the $\tau$-lepton decay products. 
The acoplanarity angle $\Phi^\ast$, defined between the $\tau$ decay planes, is sensitive to these transverse spin correlations and is influenced by the CP-mixing angle of the Yukawa coupling. The $\Phi^\ast$ angle is directly connected to $\theta_\tau$ in the $h \to \tau \tau$ differential decay rate, with the relationship taking the form of a first-order trigonometric polynomial in $\theta_\tau$. The differential decay rate can be obtained as \cite{Berge:2014wta, Kramer:1993jn}
\begin{eqnarray}
{d\Gamma_{h\to\tau^-\tau^+\to a^- a^{'+}}}\propto 
\left(1-\frac{\pi^2}{16}b(E_+)b(E_-)\cos\left(\Phi^\ast-2\theta_\tau \right)\right)\,,
\end{eqnarray}
where $a$ and $a'$ are the tau decay products and $b(E_\pm)$ 
is the spectral functions defined in \cite{Berge:2011ij}. 

The reconstruction of $\Phi^\ast$ requires the reconstruction of the $\tau$ decay planes, which is challenging at the detector level due to the missing energy associated with $\tau$ neutrinos.  
Various methods have been developed to approximate the acoplanarity angle based on differential techniques \cite{Bower:2002zx,Desch:2003rw,Berge:2008wi,Berge:2011ij,Berge:2013jra,Berge:2014sra,Berge:2015nua}. These methods are tailored to analyze specific $\tau$ lepton decay modes and are adjusted according to the number of visible particles within the $\tau_{\rm jet}$. Among them, we consider two methods for reconstructing $\Phi^\ast$\cite{ATLAS:2022akr}. 
For the $\tau^\pm\to \pi^\pm \nu_\tau$ decay, only one visible charged particle is present. In this case, the $\tau$ plane can be reconstructed using the IP method 
based on the impact parameter of the charged pions from the two $\tau$.  On the other hand, if $\tau$ decays to 
$\rho^\pm \nu_\tau$ and $a^\pm_1\nu_\tau$,  both charged and neutral pions are present. In this case we use  the transverse momenta of the visible particles to estimate $\Phi^\ast$. 

There is a significant advantage in having two visible particles from each $\tau$ decay. Although the momentum and plane of the $\tau$ lepton decays cannot be fully reconstructed due to the missing energy from the associated neutrino, the momenta of the charged and neutral pions can still be utilized. This allows for retaining information about the $\tau$ polarization and reconstructing $\Phi^\ast$. The acoplanarity angle can be derived from the charged and neutral pions at the LHC as  \cite{Berge:2015nua,Han:2016bvf}
\begin{equation}
    \Phi^\ast = \text{arccos}\left( \hat{p}^{0+}_\perp \cdot \hat{p}^{0-}_\perp \right) \times \text{sgn}\left(\hat{p}^-\cdot \left( \hat{p}^{0+}_\perp\times \hat{p}^{0-}_\perp\right)  \right)\,,
\end{equation}
where $\hat{p}^{\pm}$ is the unit vector of the charged pion three momentum in the zero momentum frame of the $\rho$-meson pair,
and $\hat{p}^{0+}_\perp, \hat{p}^{0-}_\perp$ are normalised three momentum vectors of neutral pions transverse to the charged pion momentum. 
Another requirement is the discrimination of phase space with different $\tau$ polarization. The sign of the product of the $\tau$ lepton spin analyzing function, $Y = y_-^\rho\times y_+^\rho$ , where $y_\pm^\rho = \frac{E_{\pi^\pm}-E_{\pi^0}}{E_{\pi^\pm}+E_{\pi^0}}$, and $E_{\pi}$ is the pion energy in the laboratory frame, appears in the CP-mixing sensitive terms of the squared matrix element.
As $Y$ is not positive definite,  integration over pion momenta for both $Y>0$ and $Y<0$ would average out the CP mixing sensitive terms in the matrix element.  
Accordingly, the events from different classes are separated by shifting the events with $Y<0$ by $\pi$. This way the acoplanarity angle is modified for the case of $Y<0$ only  and defined as 
\begin{equation}
    \Phi^\ast = \begin{cases}
        & \Phi^\ast-\pi \hspace{6mm} \text{if  }\hspace{4mm} 0 < \Phi^\ast<\pi \,,\\ 
        & \Phi^\ast+\pi \hspace{6mm} \text{if  }\hspace{4mm} -\pi < \Phi^\ast<0  \,.
    \end{cases}
    \label{eq:4}
\end{equation}
With  this definition $\Phi^\ast$ has the range of  $-\pi<\Phi^\ast<\pi$. 


Analogously to the $\rho$ decay mode, the acoplanarity angle of the $\tau$ decay mode $\tau^\pm\to a^\pm_1 \nu_\tau\to\pi^\pm 2\pi^0\nu_\tau$ can be constructed by considering the four momenta sum of the neutral pions as taken in the neutral component of the $\rho$ method.

For the decay mode $\tau^\pm \to \pi^\pm \nu_\tau$, the IP method is employed to reconstruct $\Phi^\ast$. The impact parameter is defined as the shortest distance between the primary vertex and the pion momentum vector extended in the direction of the $\tau$ decay point. Since it is practically impossible to reconstruct the $\tau$ lepton momentum due to the presence of $\tau$ neutrinos among the decay products, the $\tau$ lepton decay plane is reconstructed from the track momentum and the impact parameter of the charged pion. First, the normalized impact parameters of the charged pions, $\hat{n}^\ast = (0,\vec{n}^{\ast\pm})$, are measured in the lab frame and then boosted to the zero momentum frame of the visible $\pi^\pm$ pair. 
The transverse components of the boosted impact parameters to the direction of the associated charged pion momentum, $\hat{n}^{\ast\pm}_\perp$, are used to define the acoplanarity angle as follows:
\begin{equation}
    \Phi^\ast = \text{arccos} \left(\hat{n}^{\ast +}_\perp\cdot \hat{n}^{\ast -}_\perp \right)\times \text{sgn} \left(\hat{p}^{\ast -} \left(\hat{n}^{\ast +}_\perp\times \hat{n}^{\ast -}_\perp \right)  \right) \,.
\end{equation}

Considering this setup, we can analyze the CP properties of Higgs boson decays to $\tau$ lepton pairs from three hadronic decay modes of the $\tau$ lepton. In general, the IP method, which requires only one charged pion to reconstruct $\Phi^\ast$, is suitable for analyzing other decay modes such as $\tau^\pm \to \rho^\pm \nu_\tau$ and $\tau^\pm \to a^\pm_1 \nu_\tau$. However, this method has low efficiency due to the significant uncertainty associated with IP reconstruction. The IP of the  $\tau_{\rm jet}$  is relatively small compared to the tracking resolution, limiting the precision of its measurement despite the excellent resolution of the detector tracker.
An advantage of the neutral pion method is that it does not rely on the reconstruction of the IP. Instead, it requires determining the direction of the neutral pion. The relatively large distance between the primary interaction point and the electro-magnetic calorimeter (ECAL) ($\mathcal{O}(1)$ m), coupled with the fine ECAL granularity, allows the direction of neutral pions to be reconstructed with smaller relative uncertainties compared to the IP.
\section{Analysis methodology}
\label{sec:3}
With the theoretical framework established, we now proceed with a phenomenological investigation into the CP properties of the Higgs boson through its decay into a pair of $\tau$ leptons. Our analysis centers on the Higgs boson, with a branching ratio BR$(h \to \tau\tau) \sim 6.23\%$,  produced via gluon-gluon fusion with a production cross section of $51$ pb at $\sqrt{s} = 14 $ TeV and $46$ pb at $\sqrt{s} = 13.5 $ TeV. We consider the hadronic decays of the $\tau$ lepton through three distinct modes: 1) both $\tau$'s decay $\tau^\pm \to \pi^\pm \nu_\tau$, 2) both $\tau$'s decay  to $\tau^\pm \to \rho^\pm \nu_\tau$, and  3) both $\tau$'s decay to $\tau^\pm (\to a_1^\pm \nu_\tau)\rightarrow \pi^+ 2\pi^0\nu_{\tau}$, with corresponding branching fractions of a single $\tau$ lepton is  to approximately ${\rm Br}(\tau^-\rightarrow \pi^- \nu) =10.8\%$, ${\rm Br}(\tau^-\rightarrow \rho^- \nu) =25.49\%$, and ${\rm Br}(\tau^-\rightarrow \pi^- 2 \pi^0\nu)=9.26\%$ according to Ref. \cite{PDG}. 
Our networks analyse these three modes as a single input without changing network structures, and extending the analysis into mixed final states, such as $\pi \rho$, $\pi a_1$, $\rho a_1$, is straightforward. 

In this section, we discuss the construction of the combined signal and background events across different $\tau$ decay modes. Furthermore, we describe 
simulation tools we used for the simulation preserving $\tau$ spin correlation in the final states for this analysis.
\subsection{Signal reconstruction and background estimation}
ATLAS \cite{ATLAS:2022akr} and CMS \cite{CMS:2021sdq} have established the kinematic 
 selection criteria for the  CP mixing Higgs decay search in  $h\to\tau\tau$ channel. 
 We follow the selection criteria of the ATLAS analysis in this paper. 
In our simulation, tau jets are reconstructed at the detector simulation level of  Delphes \cite{deFavereau:2013fsa} 
with flat identification efficiency of $60\%$ and $1\%$ faking efficiency from light jets.
Two reconstructed $\tau$ tagged jets are required to fulfil the basic selection cuts of $P_T>20$ GeV, each containing at least one charged track with $P_T>1$ GeV inside the jet cone. Moreover, we require the reconstructed missing energy to be $\met \ge 20$ GeV.  Depending on the construction method of the decay mode, different additional selection criteria are applied to enhance the sensitivity.  

The IP method is used to reconstruct $\Phi^\ast$ for $\tau \to \pi \nu_\tau$ decays. For these events only one charged track is required for a tau decay.  The impact parameter method is applicable when IP is larger than detector resolution, therefore reconstructed $\Phi^\ast$ and CP mixing values are diluted. 
The transverse and longitudinal impact parameters $d_0$ and $z_0$ of a charged-particle track are defined as the closest distance from the primary vertex to the track in the transverse plane.  To improve the efficiency of the IP method, selected events have to satisfy $d_0 \ge 100\ \mu m$ and $z_0 \ge 2$ mm.

For events with $\rho$ meson decays a charged pion track and reconstructed $\pi^0$ are required. 
Similarly, for events with $a^\pm_1$  decays a charged pion track and two $\pi^0$ are required. 

The condition of reconstruction of $\pi^0$ in $\tau$ decays in ATLAS and CMS experiments 
are detailed in \cite{ATLAS:2015boj,CMS:2015pac,CMS:2022prd}. We imitate this condition by requiring two (four) reconstructed photons in the Delphes simulation with $P_T> 1 $ GeV and $\Delta R > 0.05$  following the suggestion in \cite{Han:2016bvf} for the $\rho^\pm $ ($a^\pm$ ) final state, respectively. To apply this condition, we use the truth information of $\tau$ jet matching to the $\rho$ ($a_1$) momentum direction.  Note that the purpose of our paper is to show the improvement using our network from DNN and not to estimate the improvement from the actual experimental situation.

For background estimation, we follow an ATLAS search for Higgs boson decaying to $\tau$ lepton pairs \cite{ATLAS:2018ynr}. The dominant irreducible background emerges from the Drell–Yan process $pp\to\gamma^\ast/Z\to\tau\tau$ which contributes $90\%$ of the total background events. Other backgrounds stem from $\bar{t}t$ and misidentified $\tau$. They are reducible and can be easily separated from two $\tau$ jet productions. Misidentifying $\tau$ rates range between $0.15-0.25$ for the one prong $\tau_{\rm jet}$ and between $0.01-0.04$ for the three prong $\tau_{\rm jet}$ \cite{ATLAS:2018ynr}. 

\begin{figure}[h!]
    \centering
    \includegraphics[scale=0.4]{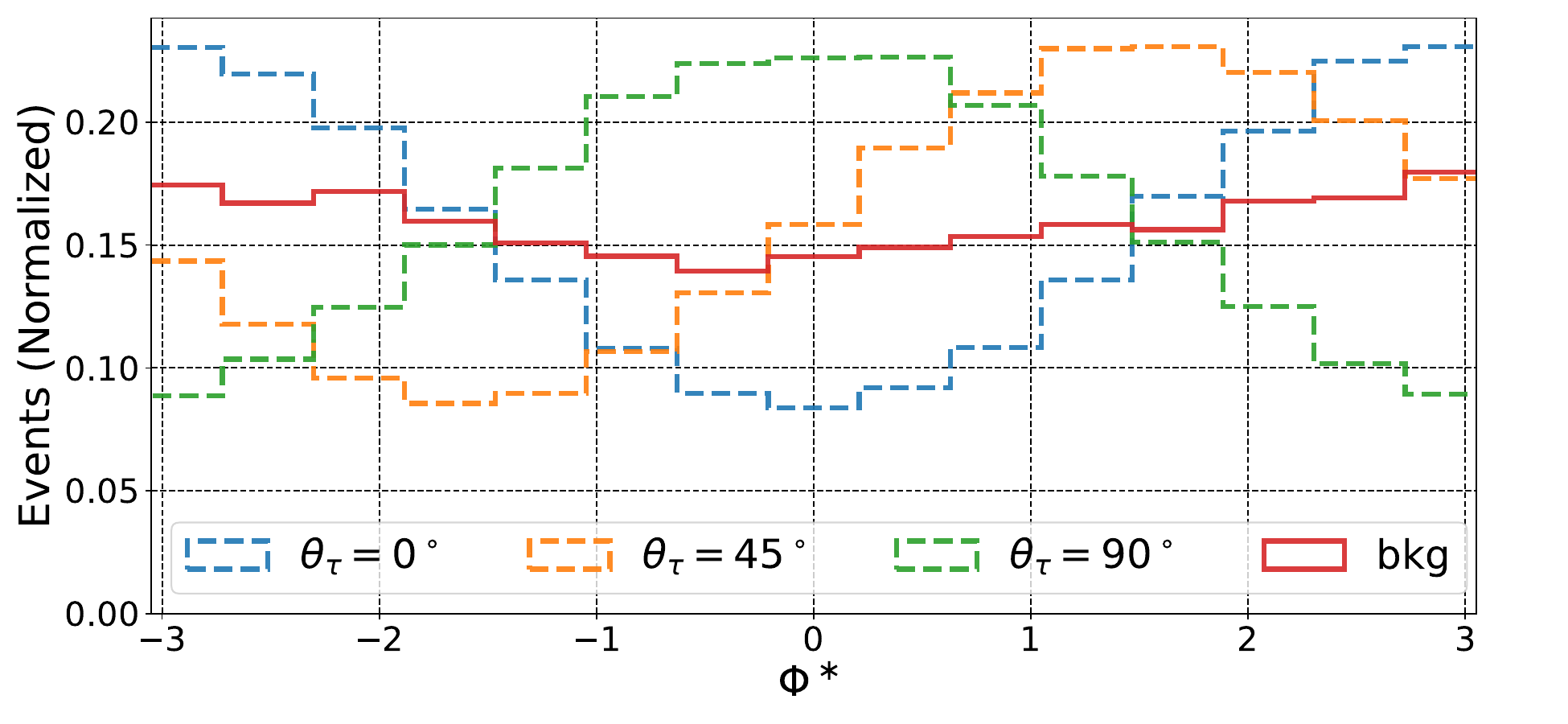}
    \caption{Normalized distributions of the reconstructed $\Phi^\ast$ at the detector level are shown for background events in red and for signal with three different CP mixing angles. The pure CP-even distribution is represented by a dashed blue line, the pure CP-odd distribution by a dashed green line, and the maximally CP-mixed state by a dashed orange line.}
    \label{fig:1}
\end{figure}

Figure \ref{fig:1}, shows the normalized distribution of the reconstructed $\Phi^\ast$ for signal events with different values of the CP mixing parameter $\theta_{\tau}$, as well as background events. Signal distributions follow the analytical distribution of the $\Phi^\ast$  as described in \cite{Berge:2015nua, Bower:2002zx}
\begin{equation}
    \alpha-\beta \cos\left(\Phi^\ast + 2\theta_\tau \right)\,,
\end{equation}
where $\alpha$ corresponds to the total cross section and $\beta$ determines the relative magnitude of the asymmetry. As clearly seen, the maximal mixing distribution with $\theta_\tau = 45^\circ$ is shifted by a phase of $2\theta_\tau$ from the pure CP-even distribution. This shift allows for effective discrimination of the CP mixing states at the reconstruction level. Such a clear reconstruction of $\Phi^\ast$ after accounting for detector effects is expected, as the impact of the detector on the neutral pion energy resolution and the charged pion transverse momentum resolution does not significantly affect the reconstructed $\Phi^\ast$ distribution. 

An important effect arises from the granularity of the ECAL in the $\eta-\phi$ plane, which impacts the angular momentum resolution in the direction of the neutral pion. This resolution is crucial for distinguishing between single photon showers and two photons from $\pi^0$ decays. This effect can obscure the differences between the reconstructed $\Phi^\ast$ distributions for various CP mixing angles. However, as shown in \cite{Han:2016bvf}, different granularity values do not significantly alter the reconstructed $\Phi^\ast$ distribution. Additionally, the positions of the minima and maxima in distributions with different CP mixing angle values remain unchanged.

\subsection{Events generation}
For event simulation, we implement the effective Lagrangian for Higgs boson production from gluon-gluon fusion using \textsc{FeynRules} \cite{Alloul:2013bka}. The NLO corrections to Higgs production are implemented as detailed in \cite{Belyaev:2012qa} giving a production cross section of $51$ pb at the energy of the HL-LHC $\sqrt{s}=14$~TeV. To incorporate CP-mixing parameters the effective coupling of the Higgs to a $\tau$ lepton pair described by Equation \ref{eq:1} is implemented into the same model files with $\kappa_\tau = 1$.

We employ \textsc{MadGraph5} \cite{Alwall:2014hca,Frederix:2018nkq} for cross section estimation and generating parton-level events.  \textsc{Pythia8.3} \cite{Bierlich:2022pfr} is utilized to include parton showering and hadronization effects. To maintain spin correlations in $\tau$ lepton decays within the matrix element, we use the \textsc{TauDecay} module \cite{Hagiwara:2012vz}, which is a part of MadGraph package. This module, integrated into the taudecayUFO model files, ensures spin correlation preservation by extending the matrix element to $2 \to N$, where $N$ represents the number of final-state pions. The factorization and renormalization scales have been kept at the default MadGraph event by event dynamic choice. Jets are formed using \textsc{FastJet} package \cite{Cacciari:2011ma} utilizing anti-KT algorithm \cite{Cacciari:2008gp} with $R=0.4$. Detector effects are taken into account with the \textsc{Delphes} package \cite{deFavereau:2013fsa} using the default ATLAS card. Three datasets for different decay modes of $\tau$ lepton are generated separately and combined with ratios according to the branching fraction of each decay mode.

For the deep learning analysis, we use \textsc{PyTorch Geometric} \cite{fey2019fast} for building the GNN networks, while standard \textsc{PyTorch} \cite{paszke2019pytorch} is used for the MLP. Finally, the  \textsc{Scikit-Learn} package \cite{pedregosa2011scikit} is used to facilitate network training and evaluation. 
\section{Deep Learning analysis}
\label{sec:4}
In this section, we explore the application of various deep learning techniques, including MLP, GCN, and GTN, to investigate the CP properties of the Higgs boson. Each network is designed to handle specific types of input data according to its structure. The MLP is adept at analyzing high-level kinematic distributions, while the GCN and GTN are suited for analyzing heterogeneous graphs constructed from final state particles.

For the MLP study, the kinematical variables  of the final state $\tau_{\rm jet}$ pairs are reconstructed, and 
$\Phi^\ast$ is calculated for the final state $\tau_{\rm jet}$ pair. The kinematical variables and $\Phi^\ast$ are fed into the MLP. Of course, the kinematical data is normalized between the standard 0 and 1 range to ensure effective processing by the neural network.  The network is trained to distinguish the background and signal processes. 
Another method for training a network to distinguish between different CP states is to use a conditional DNN network, as described in \cite{Esmail:2024gdc}.

GNNs, on the other hand, analyze graph-like structures. The standard way is that the nodes of the graph represent the final state particles, and all nodes are fully connected. The graph nodes are weighted with the four-momenta of the final state particles and edges are weighted with the angular distance between each node pair. With this approach it is not easy to incorporate the CP properties of the Higgs boson into a fully connected graph. 
Instead, we utilize a heterogeneous graph in this paper. 
A heterogeneous graph comprises multiple types of nodes and edges, each representing different entities and interactions within the experimental setup. Nodes in these graphs can still represent final state particles, but also 
the reconstructed $\tau_{\rm jet}$ and  Higgs boson. Each node type has distinct attributes and properties that define its role within the graph. Edges represent the interactions or relationships between these entities.
This method has enhanced flexibility representing the physics we focus on.

To efficiently incorporate CP information into a graph-like structure, we consider a fully connected graph of the final state pions with additional heterogeneous nodes representing the reconstructed $\tau_{\rm jet}$ and the  Higgs boson. The edges of the fully connected pion graph are weighted with the angular distance between each node pair, while the edge between the two $\tau_{\rm jet}$ nodes is weighted with the value of the reconstructed $\Phi^\ast$. By selectively connecting $\tau_{\rm jet}$ nodes to their decaying pions, we construct a graph that integrates both the kinematic information of the signal and the CP properties of different Higgs boson states.

\subsection{Multi-Layers perceptron}

 An MLP is a type of feed-forward neural network consisting of an input layer, one or more hidden layers, and an output layer. The input layer size corresponds to the number of kinematic variables being considered. Hidden layers, which contain a certain number of neurons, are where the model learns to capture the complex relationships in the data. The number of hidden layers and neurons in each layer are hyper-parameters that need to be optimized. 

\begin{figure}[ht!]
    \includegraphics[scale=0.3]{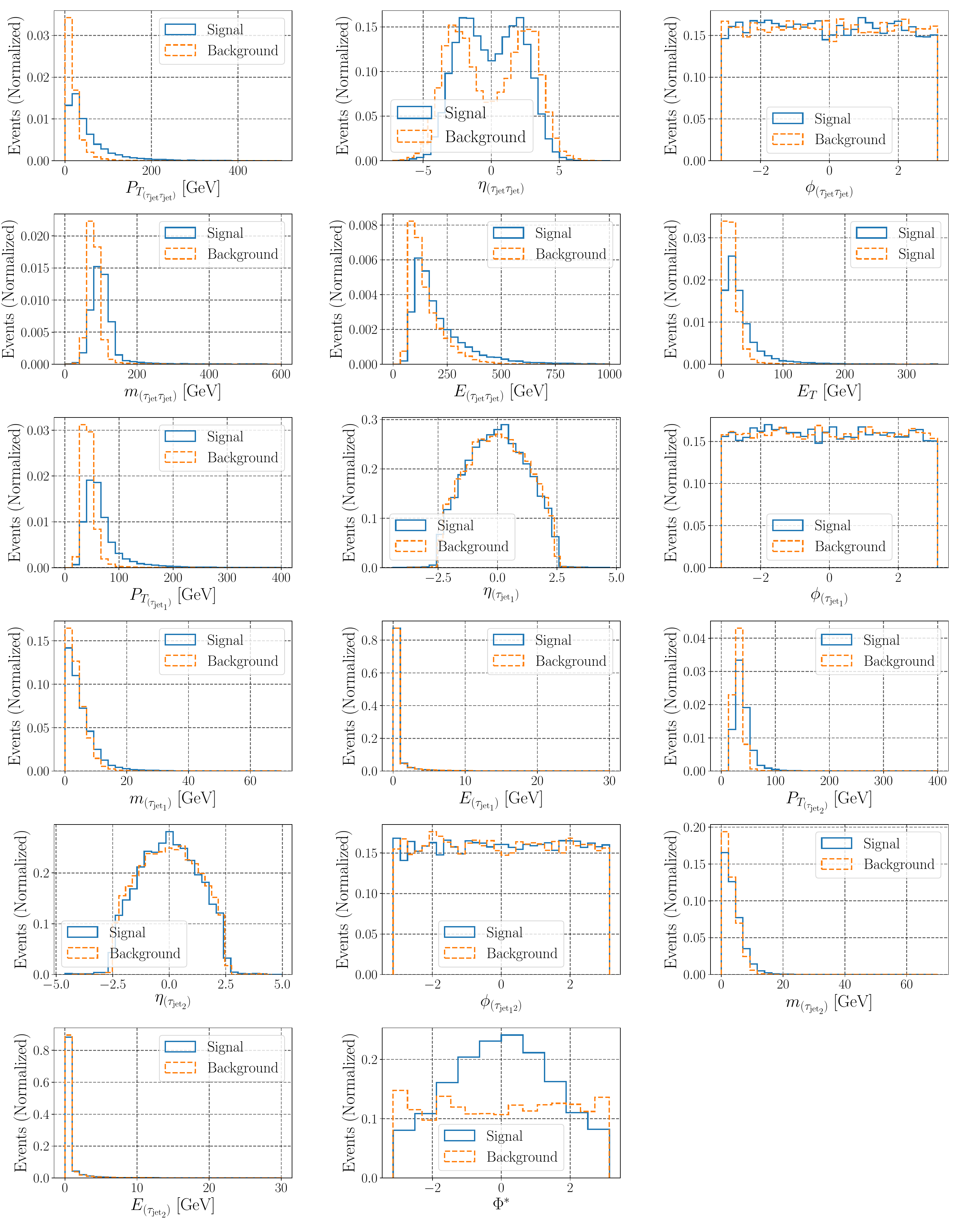}
    \caption{Kinematic distributions before applying selection cuts, used to train the MLP network for a benchmark point with $\theta_\tau = 90^\circ$. }
    \label{fig:2}
\end{figure}

The input of MLP consists of 17 inputs that encompass the kinematic and CP properties of the Higgs boson. Their distributions are shown in Figure \ref{fig:2} for a benchmark point with $\theta_\tau = 90^\circ$. In addition to the transverse momentum $p_T$, pseudorapidity $\eta$, azimuthal angle $\phi$ and invariant mass of the leading and second-leading $\tau$ jets, the input distributions include the following:

\begin{itemize}

   \item  {$\met$}: Missing Transverse Energy, defined as ${\met}  =|-\sum_{v_i}\vec{p_T}(v_i)|$, which is the sum of the transverse momenta of the visible particles.

   \item {$m_{(\tau_{\rm jet},\tau_{\rm jet})}$:} The invariant mass of the $\tau$ jet pair shows a peak around the mass of the SM-like Higgs boson for signal events, while background events peak around the mass of the Z boson. This occurs because the background events are primarily dominated by the $pp\to Z\gamma^{\ast}\to \tau\tau$ process.

   \item {$p_{T_{(\tau_{\rm jet},\tau_{\rm jet})}}$:} The transverse momentum of the $\tau$ jet pair exemplifies the slight boost of the $\tau$ jet pair in signal events compared to background events.

    \item {$E_{(\tau_{\rm jet},\tau_{\rm jet})}$:} The energy of the $\tau$ jet pair.

   \item {$\eta_{(\tau_{\rm jet},\tau_{\rm jet})}$:} The pseudorapidity of the $\tau$ jet pair.

   \item {$\phi_{(\tau_{\rm jet},\tau_{\rm jet})}$:} The azimuthal angle of the $\tau$ jet pair.

   \item {$\Phi^\ast$:} The acoplanarity angle between the two $\tau$ planes.
\end{itemize}

After reconstructing the kinematic distributions we stack all background events and signal events separately, resulting in data sets with dimensions $d_{\text{distribution}} = (17, N)$, where \(N\) is the total number of training events. We use equal size training datasets of $80000$ events for signal and background, $20000$ events are kept to evaluate the network performance during the training. For the supervised classification problem, we assign a numeric label of $Y=1$ to the signal events and $Y=0$ to the background events.

Having a suitable MLP structure that can effectively analyze the input data we scan over the MLP hyper-parameters such as the number of the hidden layers, number of neurons in each layer and the initial value of the learning rate.  The 
MLP we use consists of one input layer with a dimension equal to the input dataset dimension, 
followed by three hidden layers with rectified linear activation Unit(ReLU), where the number of neurons is  $256,128,64$ for each layer, respectively. A drop out layer is inserted after each hidden layer with a dropout rate of $10\%$ of the total number of neurons of each hidden layer. A final output layer is inserted with two neurons and softmax activation that sum up the output probability to one.
Once the training process is complete, the MLP is evaluated on the independent testing set with  $50000$ events of signal and background each, providing an unbiased evaluation of the model’s performance. 
\subsection{Graph Neural Networks}

Although MLP offers a straightforward way to analyze the CP states of the Higgs boson, it suffers from low identification performance. This is because each node in the MLP is fully connected to all other nodes in the hidden layer, which dilutes the learned CP state patterns by fully connecting the $\Phi^\ast$ distribution to the kinematic distributions. GNNs overcome this issue by analyzing heterogeneous graphs which incorporate nodes containing different types of information and selectively connected edges. This approach is well-suited for encoding kinematic and CP information.

\subsubsection{Heterogeneous Graph construction}
As mentioned previously,  we utilize a heterogeneous graph, rather than a traditional homogeneous fully connected graph, to capture the comprehensive event characteristics from the final state particles. The primary difference between heterogeneous and homogeneous graphs is that heterogeneous graphs can represent multiple types of nodes and edges, each with different properties and relationships, whereas homogeneous graphs consist of a single type of nodes and edges.
The heterogeneous graphs can  
accurately model complex systems with diverse entities and interactions, such as the decay topology of Higgs boson events. 

\begin{figure}[h!]
    \centering
    \includegraphics[scale=0.35]{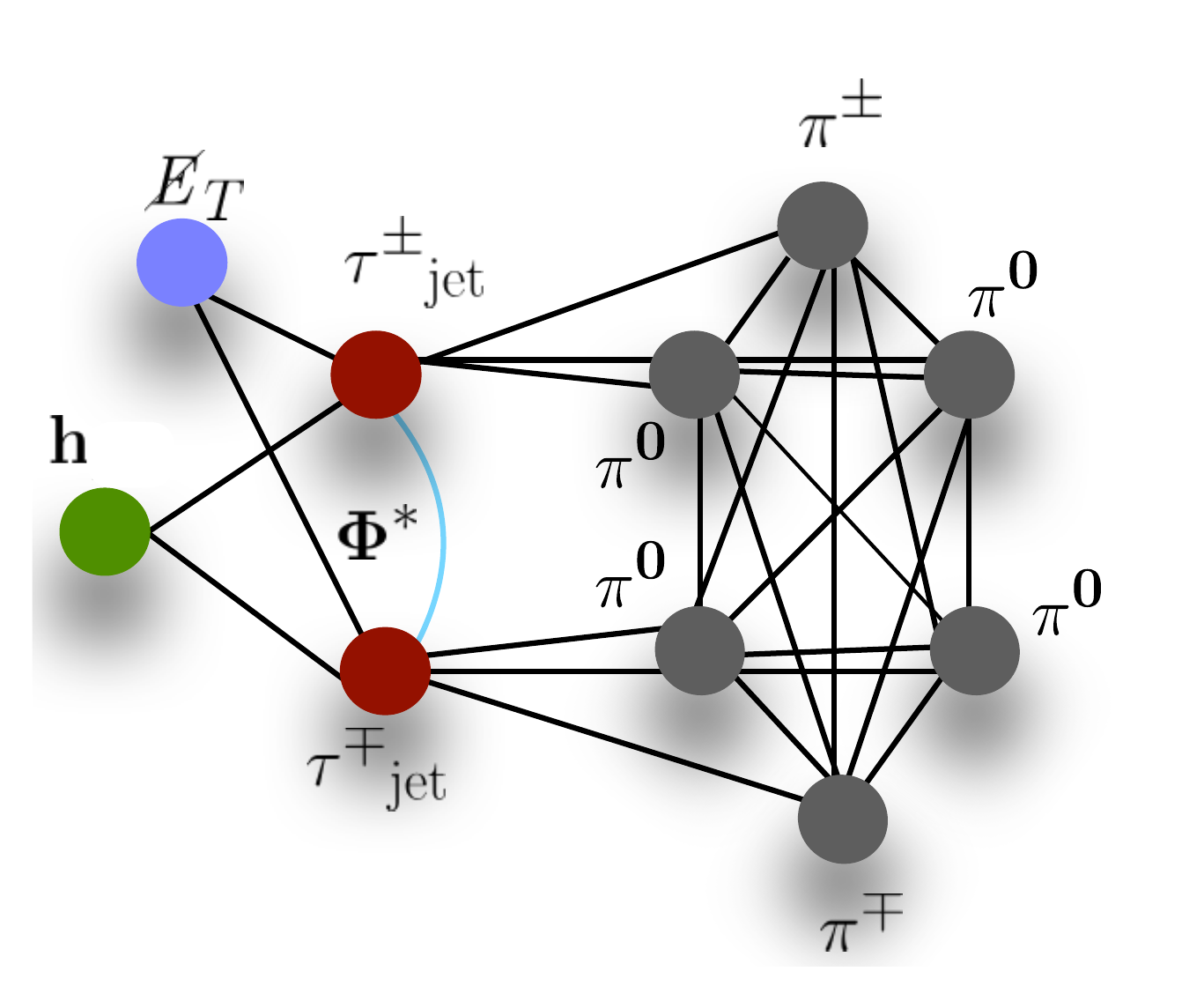}
    \caption{The constructed heterogeneous graph. Four node types are considered with different colours. Pion nodes are fully connected while other nodes are connected selectively. Edge connection between the $\tau$ jets (blue edge) are weighted with the value of the reconstructed $\Phi^\ast$.}
    \label{fig:4}
\end{figure}
In this study, we construct a heterogeneous graph from the final state particles and the reconstructed decayed particles $\tau_{\rm jet}$ and $h$. We define four node types, each representing a different type of particle and weighted with different information to encode the physical properties of the event. The first node type comprises the charged and neutral pions. We include six pion nodes for each event, representing the decay products of the $\tau$ leptons, 
up to the 3-prong decay of the $\tau$. For events with a lower number of pions, we pad the remaining nodes with zeros. The features of the pion nodes include pseudorapidity $\eta$, azimuthal angle $\phi$, angular distance from the $\tau$ jet axis $\theta_j$, transverse momentum $k_T$, the logarithm of the transverse momentum ratio of each $\tau$ constituent to the $\tau$ jet $z\equiv \log(P_T/P_J)$, and the pion energy $E$. The second node type represents the $\tau$ jets, with each event including two nodes. Their features are transverse momentum, pseudorapidity, azimuthal angle, the invariant mass and energy of the reconstructed leading $\tau$ jet. 
To encode the full event information, we introduce a third node type representing the missing transverse energy, with one node per event. This node has a single feature: the value of the missing energy in the event. 
The fourth and final node type represents the Higgs boson, with one node per event. The features of the Higgs node include pseudorapidity, azimuthal angle, transverse momentum, the invariant mass of the system of two $\tau$ jet, and their energy. Table \ref{tab:graph}  summarizes the nodes and edge features used for the heterogeneous graph construction.

\begin{table}[t]
\begin{center}
\begin{tabular}{cc}
\hline
Node name & Features \cr
\hline 
$\pi^{\pm,0}$  & $\eta_{\pi}$, $\phi_\pi$ 
$P_{T_\pi}$, $\theta_{(\pi,\tau_{\rm jet})}$, 
$E_{\pi}$, $\log(P_{T_{\pi}}/P_{T_{\tau_{\rm jet}}})$\cr
$\tau^\pm$  & $\eta_{\tau_{\rm jet}}$, $\phi_{\tau_{\rm jet}}$ 
$P_{T_{\tau_{\rm jet}}}$, $m_{\tau_{\rm jet}}$, $E_{\tau_{\rm jet}}$\cr
$\met$ & $\met$ \cr
$h$  & $\eta_{h}$, $\phi_{h}$ 
$P_{T_{h}}$, $m_h$,  where $P_h\equiv P_{\tau_{j_1}}+P_{\tau_{j_2}}$\cr
\hline
\hline
Edge name & Features \cr
\hline 
$\pi_i$ - $\pi_j$  & $\Delta R_{ij}$ \cr
$\pi_i$ - $\tau_j$  &  $\log(P_{T_i}/P_{T_{\tau_j}}), \theta_j$\cr
$\met$ - $\tau$ & $\mathbb{I}$ \cr
$h$ - $\tau$ & $\log(P_{T_{\tau}}/P_{T_{h}})$ \cr
$\tau$ - $\tau$ & $\Phi^\ast$ \cr
\hline
\end{tabular}
\end{center}
\caption{Nodes and edge features of the used heterogeneous graphs.}\label{tab:graph}
\end{table}

The graph is constructed to reflect the decay topology of the Higgs boson events, as shown in Figure \ref{fig:4}, ensuring that relevant kinematic and CP information is integrated into the graph design. Accordingly,  each of the three pion nodes is connected to the corresponding tau node, representing the decay products of the $\tau$ lepton. Both tau nodes are connected to the Higgs node, representing the Higgs boson decaying into two $\tau$ leptons. Additionally, the two tau nodes are connected by an edge weighted by the value of the reconstructed $\Phi^\ast$, capturing the relative orientation and interaction between the two $\tau$ leptons. 

This graph structure ensures that the connections between nodes represent the physical interactions occurring in the event, embedding essential physics information directly into the graph. This design enhances the model's ability to learn and infer the decay kinematics and the CP properties of the Higgs boson by leveraging the intrinsic event topology. Node types can be summarized as follows:

\begin{itemize}
    \item \textbf{Pion node}: Six pion nodes are fully connected to each other and their edges are weighted with the angular distance between each pair, $\Delta R$. 
    Although charged pion has the same charge as parent $\tau$ lepton and the role in the $\Phi^{\ast}$ reconstruction is different, we require the same type of information for both charged and neutral pions in this framework. This is because the information from both charged and neutral pions is used to reconstruct the $\Phi^\ast$ before constructing the graph.
    \item \textbf{Tau node}: The $\tau$ jet pair is connected with an edge, weighted by the value of the reconstructed $\Phi^\ast$. 
    Moreover, the $\tau$ jet is connected to the constituent pions nodes 
    weighted by the ratio of the transverse momenta and the angular distance between the corresponding piton and the $\tau$ jets. 
    \item \textbf{Missing energy node:} The missing energy node is weighted with the transverse momentum of the visible particles as $\met = -\left| \sum_{{\upsilon_i} }\vec{p}_{T_{\upsilon_i}} \right|$ and connected to each tau node weighted by a unit vector. Connecting the missing energy node to the tau nodes allows the network to fully recover all information needed to reconstruct $\tau$ leptons.  
    \item \textbf{Higgs node:} The Higgs node is weighted by the kinematic proprietaries of the reconstructed four momenta of the $\tau$ jet pair and connected only to the $\tau$ jet nodes with the ratio of the transverse momentum of the Higgs boson and the corresponding reconstructed $\tau$ jet. 
    We do not connect the Higgs node to the missing energy node because $\tau$ jet nodes are already connected to the missing energy node and there will be no information gained by adding this edge. 
\end{itemize}

Once the heterogeneous graphs are constructed,  we stack all backgrounds and signal
events separately and adjust labels with $Y=0$ and $Y=1$ for the backgrounds and signal events, respectively. During the training process, the model tries to minimize the difference between its predictions and the assigned labels using cross-entropy loss as  $-\sum Y(x)\log(\hat{Y}(x))$, with $Y$ and $\hat{Y}$ are the true and predicted labels for each class.  

\subsubsection{GNN training  on heterogeneous graphs}
Heterogeneous graphs contain various types of information linked to their nodes and edges, making it impossible for a single feature tensor to represent all node or edge features across the entire graph, due to variations in type and dimensionality. Instead, distinct types must be defined for both nodes and edges, each associated with its data tensors, as we have done in Sec 4.2.2. 
The message-passing framework is modified so that the computation message and update functions are node and edge-specific. Accordingly, training of GNNs on a heterogeneous graph is different from homogenous GNN training.  For the training process we follow the methodology introduced in \cite{schlichtkrull2018modeling} which is detailed as follows:

\begin{figure}[h!]
    \centering
    \includegraphics[scale=0.175]{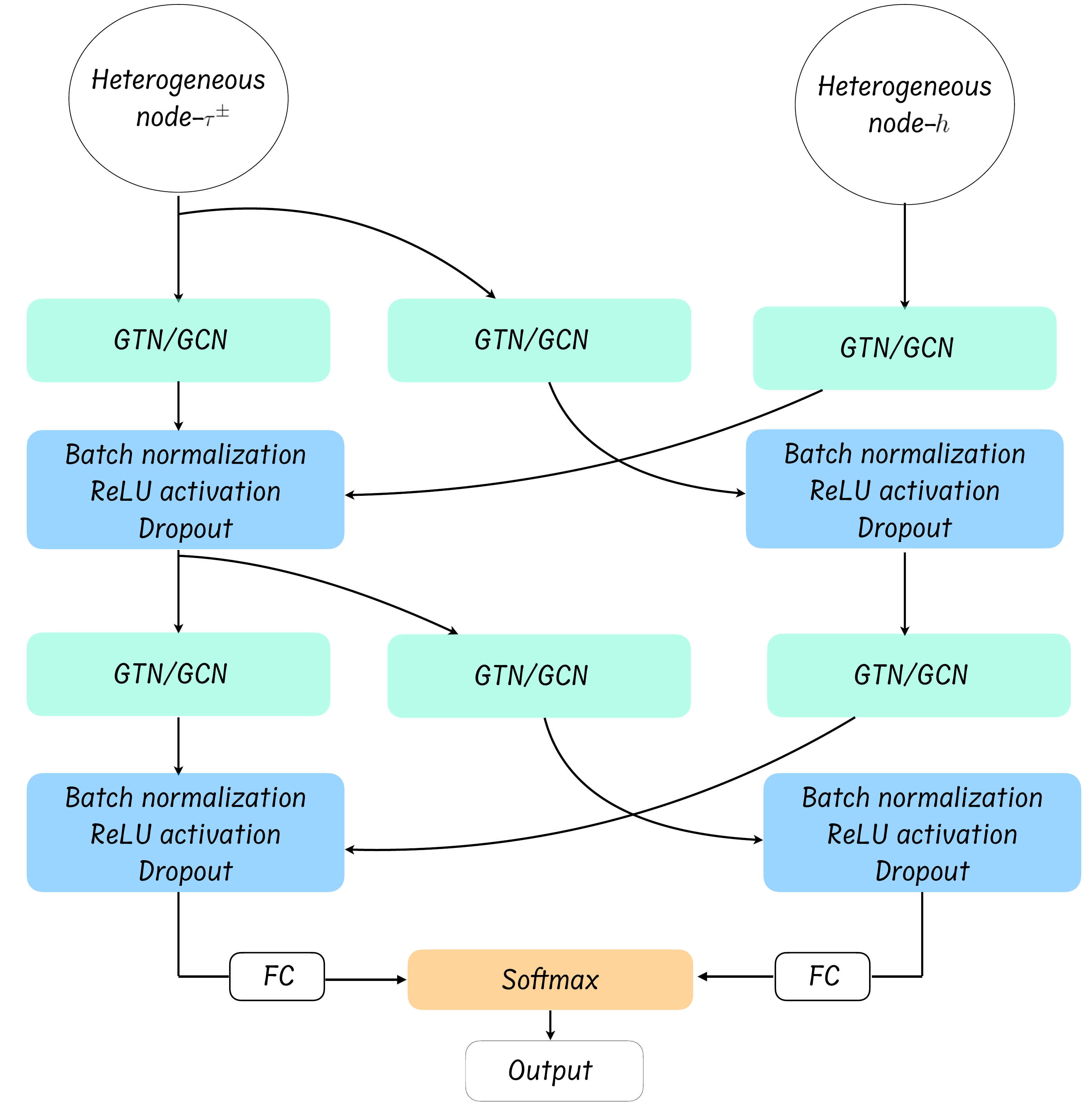}
    \caption{A schematic representation of message passing between $\tau^\pm_{\rm jet}$ and $h$. This diagram illustrates message passing between two heterogeneous nodes for illustrative purposes only; the actual network includes all the nodes shown in Figure \ref{fig:4}.
    }
    \label{fig:3}
\end{figure}

\begin{itemize}

\item \textbf{Message Passing and Node Update:}
Message passing is performed separately for each edge and node type using edge-specific convolution operations. This separation is necessary because nodes and edges containing different information cannot be processed by the same function. Consequently, each edge connection requires an additional GNN layer to adjust the output dimensions before passing information to other nodes in the graph. For example, messages passed from a $\tau^\pm_{\rm jet}$ node to an $h$ node involve three edge connections: one from $\tau^\pm_{\rm jet}$ to $h$, one from $h$ to $\tau^\pm_{\rm jet}$, and one from $\tau^\pm_{\rm jet}$ to itself (as the two $\tau^\pm_{\rm jet}$ nodes are interconnected). Each of these three edges necessitates its own GNN layer plus an additional layer, as illustrated in figure \ref{fig:3}. The additional layers are incorporated 
to have the same output dimensions for different inputs across all nodes. 
In this case, the outputs of the GNN layers are a vector of a fixed size of $32$.

\item \textbf{Aggregation and Pooling:}
Once the node embeddings for all types are updated, a global graph-level representation is needed for graph classification tasks.  Because all outputs have the same size, the outputs from the GNN layers are summed to a single vector for pooling, and the ReLU activation function is applied. 
One hidden layer comprises GNN layers, activation functions, batch normalization, and dropouts. These can be repeated multiple times to capture the complex structure of the input data.

\item \textbf{Graph-Level Classification:}
For graph classification, this typically involves fully connected layers followed by a softmax activation function that outputs two probability values indicating the class, either a signal or background-like.
It is important to note that Figure \ref{fig:3} serves solely as an illustration of how the $\tau^\pm_{\rm jet}$ and $h$ nodes are updated, whereas the actual training involves message passing for all node connections, as depicted in Figure \ref{fig:4}.
\end{itemize}

This training structure allows for capturing the graph's heterogeneous nature, ensuring that information from different node and edge types is effectively utilized in the classification task.

\subsubsection{Graph Convolution network}
GCNs have gathered significant attention in recent years for their ability to learn representations of graph-structured data\cite{Bachlechner:2022cvf,Esmail:2023axd,Sahu:2024sts}.  
The primary goal of a GCN is to learn a function that maps input features to new representations capturing the relationships among the graph nodes 
\cite{Bachlechner:2022cvf,Esmail:2023axd,Sahu:2024sts}.
The core concept behind GCN is the generalization of the convolution operation from regular grids to irregular graphs. A graph convolution operation can be seen as a local averaging of features from neighboring vertices capturing both the local structure of the graph and the features associated with each node. Given an input graph $G = (V, E)$, the graph convolution operation is defined as
\begin{equation*}
H^{(l+1)} = \sigma \left( \hat{D}^{-\frac{1}{2}} \hat{A} \hat{D}^{-\frac{1}{2}} H^{(l)} W^{(l)} \right),
\end{equation*}
where $H^{(l)} \in \mathbb{R}^{N \times F_l}$ is the feature matrix at layer $l$, with $N$ being the number of vertices in the graph and $F_l$ the dimension of the feature space at layer $l$. $W^{(l)} \in \mathbb{R}^{F_l \times F_{l+1}}$ is the learnable weight matrix at layer $l$. Furthermore, $\sigma$ denotes the activation function. The matrix $\hat{A} \in \mathbb{R}^{N \times N}$ is the adjacency matrix of the input graph with added self-connections, defined as $\hat{A} = A + I_N$, where $A$ is the adjacency matrix of $G$  and $I_N$ is the identity matrix of size $N$. The matrix $\hat{D} \in \mathbb{R}^{N \times N}$ is a diagonal matrix with $\hat{D}_{i} = \sum_j \hat{A}_{ij}$ representing the degree of vertex $i$ in the graph with added self-connections. The graph convolution operation can be interpreted as a message-passing mechanism, where each vertex aggregates information from its neighbours and updates its features according to the learned weights. This process is repeated over several layers, allowing the model to capture higher-order relationships between vertices in the graph. 
We found the optimal architecture consists of three hidden layers. 
Each hidden layer is followed by a rectified linear unit (ReLU) activation function. 
The model employs sum aggregation to consolidate information from all nodes in the graph, followed by a final classification layer. The model training process was 
managed by a learning rate scheduler with an initial learning rate of 0.001, a step size of 5 epochs and a decay factor of 0.8. 
This configuration was empirically determined to yield the best performance in our evaluations. 

Although GCN has demonstrated considerable success, it also has several key limitations. 
One major drawback of the traditional GCN model is its inability to perform inductive learning tasks due to its dependence on the graph's specific adjacency matrix. The GCN model is also constrained by its rigid neighbourhood aggregation method. Moreover, GCN uniformly weights all neighbouring nodes during feature aggregation, which can be ineffective if some neighbours carry more significant information than others, which is often the case for our heterogeneous graph structure. These challenges have inspired the creation of numerous GCN variants, such as GTN \cite{shi2020masked}.

\subsubsection{Graph Transformer Network}

GTN is an advanced approach to handling graph-structured data, leveraging the power of transformer architectures to process information. Transformers were originally developed for natural language processing and introduced in LHC analyses in \cite{Mikuni:2021pou, Kach:2022uzq,DiBello:2022iwf,Qu:2022mxj, Hammad:2023sbd,Hammad:2024cae,Hammad:2024hhm}. 
GTN combines the strengths of traditional GNN with transformers to provide a more powerful and flexible framework for analyzing graphs. The attention mechanism of the transformer enables the network to selectively focus on different nodes of the input graph, allowing for the modelling of complex relationships and dependencies. 
Therefore, GTN is suitable for extracting diverse information from nodes with different structures, e.g. pion and $\tau$ jet nodes enabling the effective extraction of kinematic and CP information about the signal events.

GTN works by firstly embedding the graph by passing the  node features $\alpha_i$, for node $i$, and edge features $\beta_{ij}$, for each edge between the nodes $i$ and $j$, by a linear projection layer  as 
\begin{align}
h_i = A\ \alpha_i + a_0\,, \hspace{8mm}
e_{ij} = B\  \beta_{ij} + b_0 \,,
\end{align}
where $A, B$ and $a_0, b_0$ are the trainable matrices and biases of the linear projection layer for the edge and node features, respectively.

After the graph embedding, the self-attention mechanism is applied. This mechanism allows each node to attend to all other nodes in the graph. The attention mechanism calculates attention scores based on the similarities between the feature vectors of nodes. These scores determine the influence of neighbouring nodes on the target node.
The attention mechanism works by defining key $K$,  query $Q$, and calculates an attention matrix $\alpha$ as follows:
\begin{equation}
\alpha_{ij} = \frac{Q\cdot K^T}{\sqrt{d_k}} \,,
\end{equation}
where $Q = W_Q\cdot h_i,K = W_K\cdot h_j$ with $W_Q$ and $W_K$ learned weight matrices, and $d_k$ is the dimension of the key vectors.
For edge-wise attention, the mechanism works as
\begin{equation}
A_{ij} = \frac{\exp(\alpha_{ij}  \cdot E^k_e)}{\sum_{k \in \mathcal{N}(i)} \exp(\alpha_{ij} \cdot E^k_e)} \,,
\end{equation}
where $\mathcal{N}(i)$ denotes the neighbours of node $i$,  $E^k_e = e_{ij}\cdot W_E$ is a learnable weight vector for edge type embeddings,  and $j$ stands for the other vertex where edge $k$ is attached.

\begin{figure}[h!]
    \centering
    \includegraphics[scale=0.35]{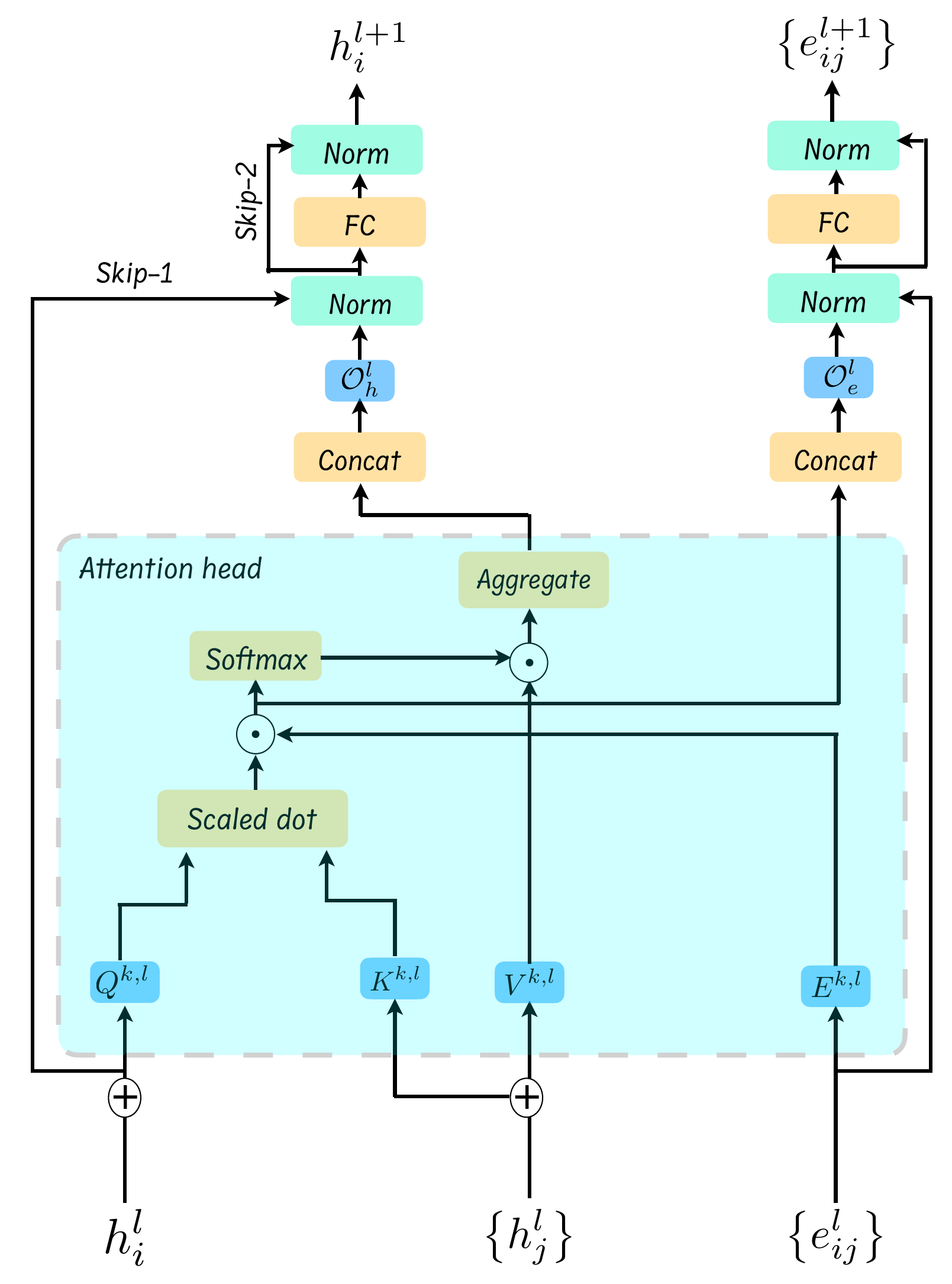}
    \caption{Schematic representation of a GTN layer. See the text for a detailed description of the GTN.  }
    \label{fig:5}
\end{figure}

The attention scores are used to aggregate information from neighbouring nodes. Each node updates its feature vector by taking a weighted sum of its neighbours’ features based on the weights derived from the attention scores. This process is analogous to how traditional GNNs aggregate information from neighboring nodes, but with the added flexibility of attention weights. The node update has the form
\begin{equation}
h_i^{(l+1)} = \sigma \left( \sum_{j \in \mathcal{N}(i)} m_{ij} + h_i^{(l)} \right) \,,
\end{equation}
where $\sigma$ is an activation function and $l$ denotes the layer number. The message passing, $m_{ij}$, is extended to include the attention effect as
\begin{equation}
   m_{ij} = A_{ij} (W_V h_j + e_{{ij}}) \,,
\end{equation}
where   $W_V$ is a learnable weight matrix. The edge update has the same form as the node update.  Note that  $W_V h_j + e_{ij}$ works as the value of the attention mechanism. 
To capture different types of relationships and interactions GTNs employ multi-head attention. Multiple attention mechanisms run in parallel, each focusing on different aspects of the node features and interactions. The results are then concatenated and linearly transformed to produce the final node embeddings. 
 The final node embeddings are processed through additional layers, a feed-forward neural network, with residual connections to produce the desired output in the form, 
\begin{equation}
h_i^{(l+1)} = \mathcal{F} \left( h_i^{(l)} \| h_j^{(l)} \| e_{ij}^{(l)} \right)
\end{equation}
where $\mathcal{F}$ is a feed forward neural network and $\|$ denotes concatenation over all parallel attention heads. A schematic representation of a GTN layer is depicted in figure \ref{fig:5}.

The optimized structure of the used GTN is determined through empirical evaluation and is comprised of four GTN layers for each type of graph node, with an additional three layers to adjust the dimensions of the different nodes and edges in the graph. All GTN layers comprise eight attention heads. The output of these layers is fixed to a vector of length $32$ and passed by the ReLU activation function to incorporate non-linearity. To enhance training stability batch normalization is applied after each ReLU activation. Dropout is incorporated to mitigate over-fitting by randomly deactivating $10\%$ of neurons during the training process. This hidden layer, GTN layers, ReLU activation, batch normalization and dropout are repeated three times. We use the sum aggregation function to integrate information from all graph nodes, leading to a final fully connected classification layer with two output neurons.  Similar to the GCN case the training was conducted with a learning rate of $0.001$, managed by a scheduler with a step size of $5$ epochs and a decay factor of $0.8$.

\section{Results}
\label{sec:5}

\begin{figure}[h!]
    \centering
    \includegraphics[scale=0.45]{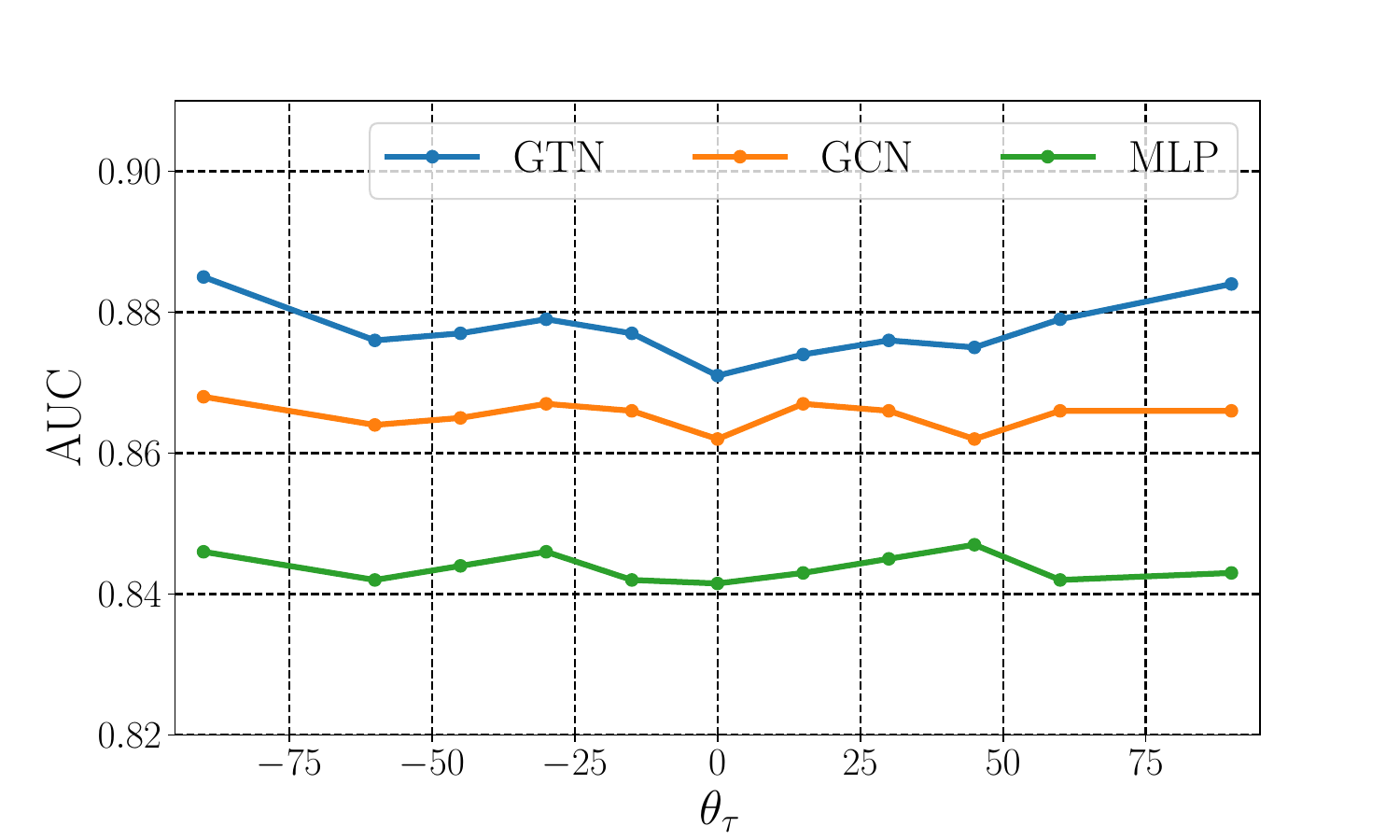}
    \caption{Area Under the ROC Curve (AUC) values for the three networks represented  by the filled bullets for eleven CP mixing angles
    $\theta_\tau$ ranging from $-90^\circ \ \rm to \ 90^\circ$. Each network was trained and tested using the signal samples of eleven individual $\theta_\tau$ values.
    }
    \label{fig:6}
\end{figure}

The discrimination power of each network is measured by the background rejection for a given signal efficiency. 
The discrimination power is intertwined with various kinematic distributions, including CP information.  
We utilize three different neural networks: MLP, GCN, and GTN, each trained on eleven benchmark points with $\theta_\tau$ ranging from $-90^\circ \  \rm to \ 90^\circ$. Each network was trained and tested on each benchmark point individually.
We use the area under curve (AUC) of the receiver operating characteristic curve (ROC) to assess the networks' performance. ROC is a curve of the True Positive Rate (TPR) as the function of the False Positive rate (FPR), and AUC is the area surrounded by  ${\rm FPR}=1$ and ${\rm TPR}>0$. 
Figure \ref{fig:6} shows the AUC values for all eleven signal points for MLP (green), GCN (orange), and GTN (blue). GTN demonstrates superior performance with an AUC of approximately $88\%$ for all points, while  GCN and MLP achieved AUCs of approximately $86\%$ and $84\%$, respectively.
Note that the AUC does not depend on $\theta_{\tau}$ so that the use of DL does not introduce additional bias to the analysis.

We also list the number of misclassified events for $y>y_{\rm cut}$ where signal efficiency is 80\% in table \ref{tab:1} for a benchmark point with $\theta_\tau=90^\circ$ for integrated luminosity of $100$~fb$^{-1}$ at $\sqrt{s}=14$ TeV\footnote{We take the integrated luminosity close to the current LHC analysis for comparison. }.
The acceptance of the background events is 1.7\% for MLP, 1.0\% for GCN and 0.7\% for GTN. 
We also compute the signal significance, following \cite{LHCDarkMatterWorkingGroup:2018ufk,Antusch:2018bgr}
\begin{equation}
\sigma_{\rm sys}= \sqrt{2\cdot\left[(S+B)\cdot \ln\left(\frac{(S+B)(B+\delta^2_B)}{B^2 +(S+B)\delta^2_B}\right)-\frac{B^2}{\delta^2_B}\cdot \ln\left(1+\frac{\delta^2_B S}{B(B+\delta^2_B)}\right)\right]} \,,
\label{eq:sig_Z}
\end{equation}
where $S, B$ is the number of the signal and background events, and $\delta_B$ represents the systematic uncertainty of the SM background events and is set to $20\%$ \cite{ATLAS:2022akr}. These results demonstrate that the $ h \to \tau\tau $ process can be effectively identified at the HL-LHC by using the proposed networks.

\begin{table}[h!]
\centering
\resizebox{\textwidth}{!}{
\begin{tabular}{lcccc}

                      &Selection cuts & MLP(TPR$>0.8$)&GCN(TPR$>0.8$)& GTN(TPR$>0.8$)\\
   
    \hline
    Background events & 872554 & $ 14982 $  &$8901$& $6169$   \\  
    \hline

    Signal events   &1102 & $703$  &$705$& $708$   \\  
    \hline

     Signal significance & $2.9\sigma$ &$5.6\sigma$  &$7.2\sigma$& $8.6\sigma$  \\  
    \hline
\end{tabular} }
\caption{Number of signal and background events at HL-LHC with energy $\sqrt{s}=14$ TeV and integrated luminosity $\mathcal{L}=100\rm fb^{-1}$ for a benchmark point with $\theta_\tau=90^\circ$. The first column displays the number of signal and background events after the selection cuts. The subsequent columns present the number of signal and background for the used DNNs with a True Positive Rate exceeding 0.8. The last row presents the signal significance, calculated using equation \ref{eq:sig_Z}.}
\label{tab:1}
\end{table}

To better understand the outputs of different networks, particularly the feature regions each network focuses on to achieve its classification performance, we use Shapley Additive Explanations (SHAP) \cite{lundberg2017unified}.
SHAP is a method to estimate an importance value for each feature using the output of deep learning models. 
For a given prediction $ f(x)$, the SHAP value for a feature $i$ is calculated as:
\begin{equation}
\phi_i = \sum_S  \frac{|S|! (|N| - |S| - 1)!}{|N|!} \left[ f(S \cup \{i\}) - f(S) \right] 
\end{equation}
where $N $ is the set of all features, $S$ is a subset of $N$  excluding feature $ i $, $ f(S)$ is the prediction based on the features in subset $S$, and $f(S \cup \{i\}) $ is the prediction with feature $ i $ added to $ S $. The SHAP value $\phi_i $, thus, represents the average contribution of feature $i $ to the prediction over all possible subsets $S$. This method ensures a fair and consistent allocation of feature importance considering the correlation between the input distributions. Since the networks used have a static structure with a fixed input dataset size, feature $ i $ is randomly sampled to eliminate its impact on the network output.

\begin{figure}[h!]
    \centering
    \includegraphics[scale=0.25]{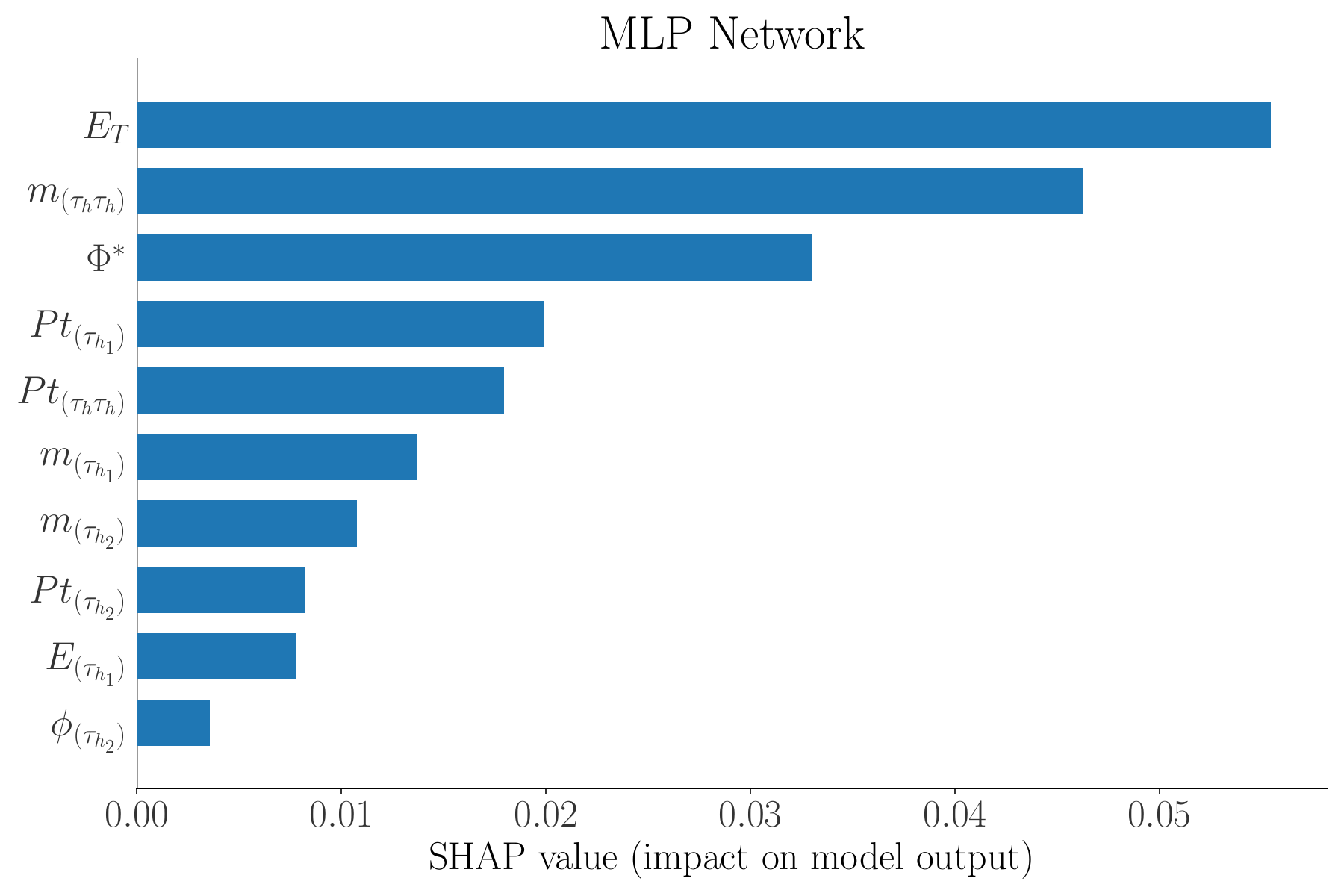}~~\includegraphics[scale=0.25]{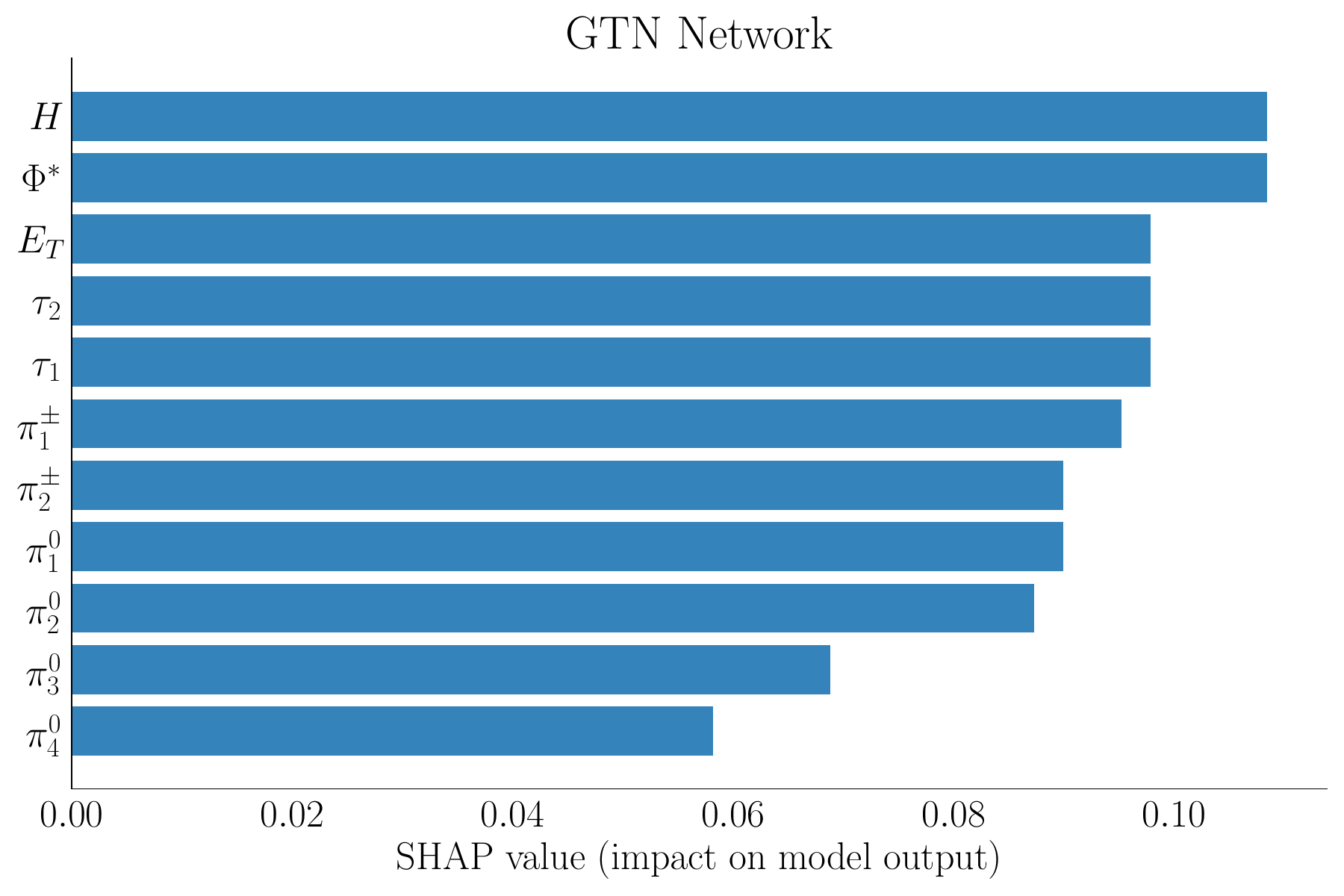}
    \caption{Average SHAP values for $10000$ test events for a signal with $\theta_\tau=90^\circ$ for MLP (left) and GTN (right). GTN plot shows the SHAP values for all the graph nodes and the $\Phi^\ast$ edge.}
    \label{fig:shap}
\end{figure}

Figure \ref{fig:shap} shows the SHAP values for 10000 test events for a signal point with $\theta_\tau=90^\circ$ for the MLP (left) and GTN (right). SHAP values are computed for the inputs of both networks, including the reconstructed distributions shown in Figure \ref{fig:2} and the graph nodes plus the $\Phi^\ast$ edge shown in Figure \ref{fig:4}. The results indicate that in both networks, $\Phi^\ast$ significantly affects the network output. Interestingly, the MLP focuses primarily on the information from $\met,\Phi^\ast, p_{T_{(\tau_h\tau_h)}}$ and the transverse momentum of the leading $\tau$ jet, while it is less sensitive to all other input distributions. Conversely, the GTN distributes its attention almost equally across all nodes in the input graph, which may explain the improved performance of the GTN over the MLP.

\subsection{Shape analysis}
\begin{figure}[h!]
    \centering
   \includegraphics[scale=0.35]{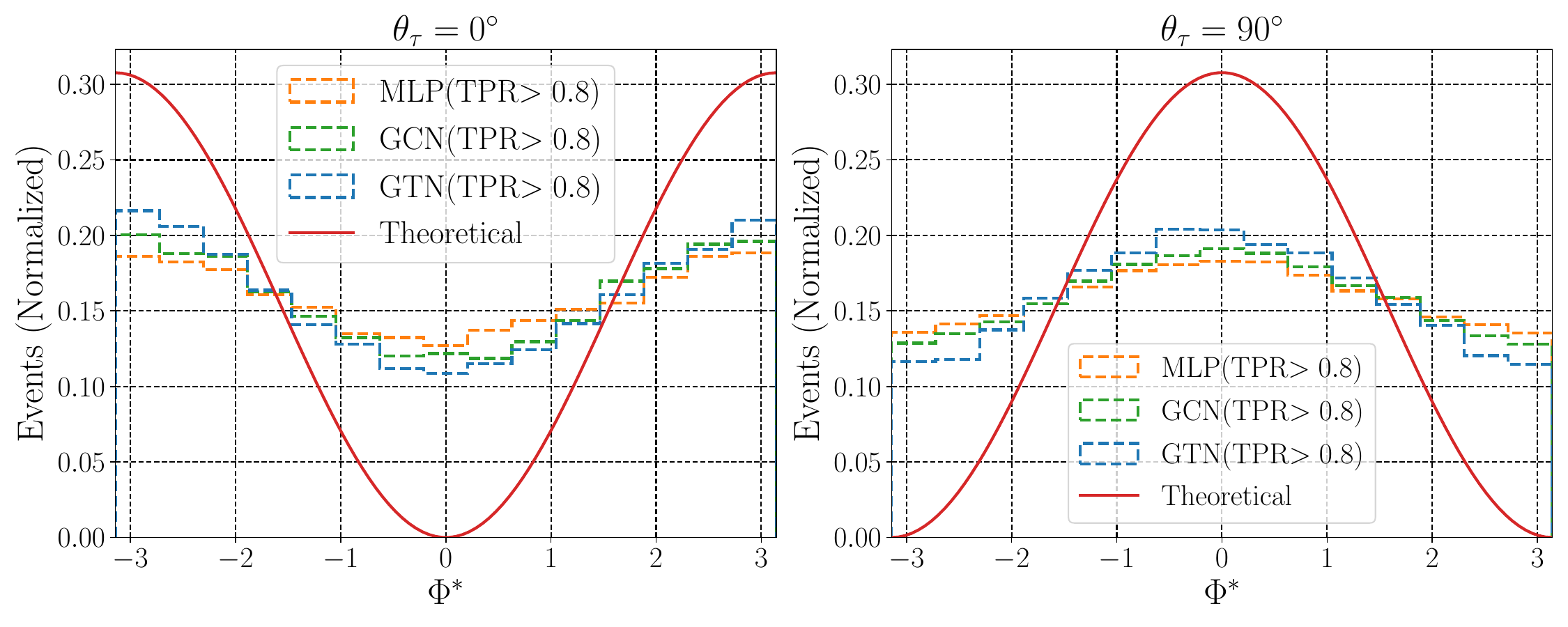}
    \caption{Acoplanarity angle distributions of the signal for CP mixing $\theta_\tau=0^\circ$ (left) and $\theta_\tau=90^\circ$ (right). Dashed histograms represent  $\Phi^\ast$ for events with True Positive Rate (TPR $ > 0.8$) for the three networks when tested on $50000$ events for signal and background each. The theoretical prediction is represented by the red solid line.}
    \label{fig:7}
\end{figure}

In this subsection, we explore the measurement of the CP state of the Higgs boson at the LHC by analyzing the $\Phi^\ast$ distribution for all considered $\theta_\tau$ values. We focus on the $\Phi^\ast$ distribution after maximizing the performance of the DNNs with a TPR $\ge 0.8$. Figure \ref{fig:7} shows normalized $\Phi^\ast$ distributions after applying a TPR cut for the three networks, considering two benchmark points with CP-mixing angles $\theta_\tau=0^\circ$ (left) and $90^\circ$ (right). These distributions are obtained from a test sample of 50000  signal and background events, but still peaks at the correct location. Note that because of the high rejection efficiency of the background, the distribution is mostly of the signal. For example, the contribution of the background is less than 1/80 of the signal for GTN.  At $\theta_{\tau}=0$, the ratios of the minimum and maximum of the distributions are 0.13/0.19=0.68 for MLP while it is 0.10/0.22 =0.45 for the GTN, showing the signal distribution is reconstructed correctly by using GTN.

We then use these distributions for a binned log-likelihood analysis to test the probability of measuring a non-zero CP mixing angle against the CP-even SM case \cite{Han:2016bvf}.
To do so, we start by constructing a likelihood function, $\mathcal{L}(D|\theta_\tau)$, which represents the probability of observing data $D$ for a given parameter $\theta_\tau$. 
For hypothesis testing the negative log-likelihood ratio compares the likelihoods of two hypotheses: the null hypothesis $\mathcal{L}(D|0)$ and the alternative hypothesis $\mathcal{L}(D|\theta_\tau)$. The null hypothesis represents the probability of observing data consistent with a purely CP-even Higgs combined with background events, while the alternative hypothesis represents the probability of observing data with $\theta_\tau \ne 0$ combined with background events. The limit on observing a non-CP-even state is determined by rejecting the null hypothesis at a certain confidence level. The binned negative log-likelihood ratio is defined as \cite{Han:2016bvf}
\begin{equation}
    -\Delta\ln\mathcal{L} = -\sum_i \left[n_i\log \left(\frac{n_i}{\nu_i} \right)+\nu_i-n_i \right]\,, 
    \label{eq:6}
\end{equation}
where the sum runs over the bins of the two hypothesis histograms, $n_i, \nu_i$. Under the null hypothesis, and for large sample size, the test statistics $ -\Delta\ln\mathcal{L}$ approximately follows a Chi-squared distribution with degrees of freedom equal to the difference in the number of parameters between the two hypotheses \footnote{In our case we consider a $\chi^2$ distribution with degree of freedom equal to one, which represents the CP-mixing angle.}. Using the $\chi^2$ distribution, the P-value can be computed as $P(\chi^2 \ge \Lambda)$, where $\Lambda$ is the test statistic value of $ -\Delta\ln\mathcal{L} $. The confidence level of rejecting the null hypothesis and obtaining a limit on the non-zero CP-mixing is $1-P\text{-value}$.

\begin{figure}[h!]
    \centering
    \includegraphics[scale=0.45]{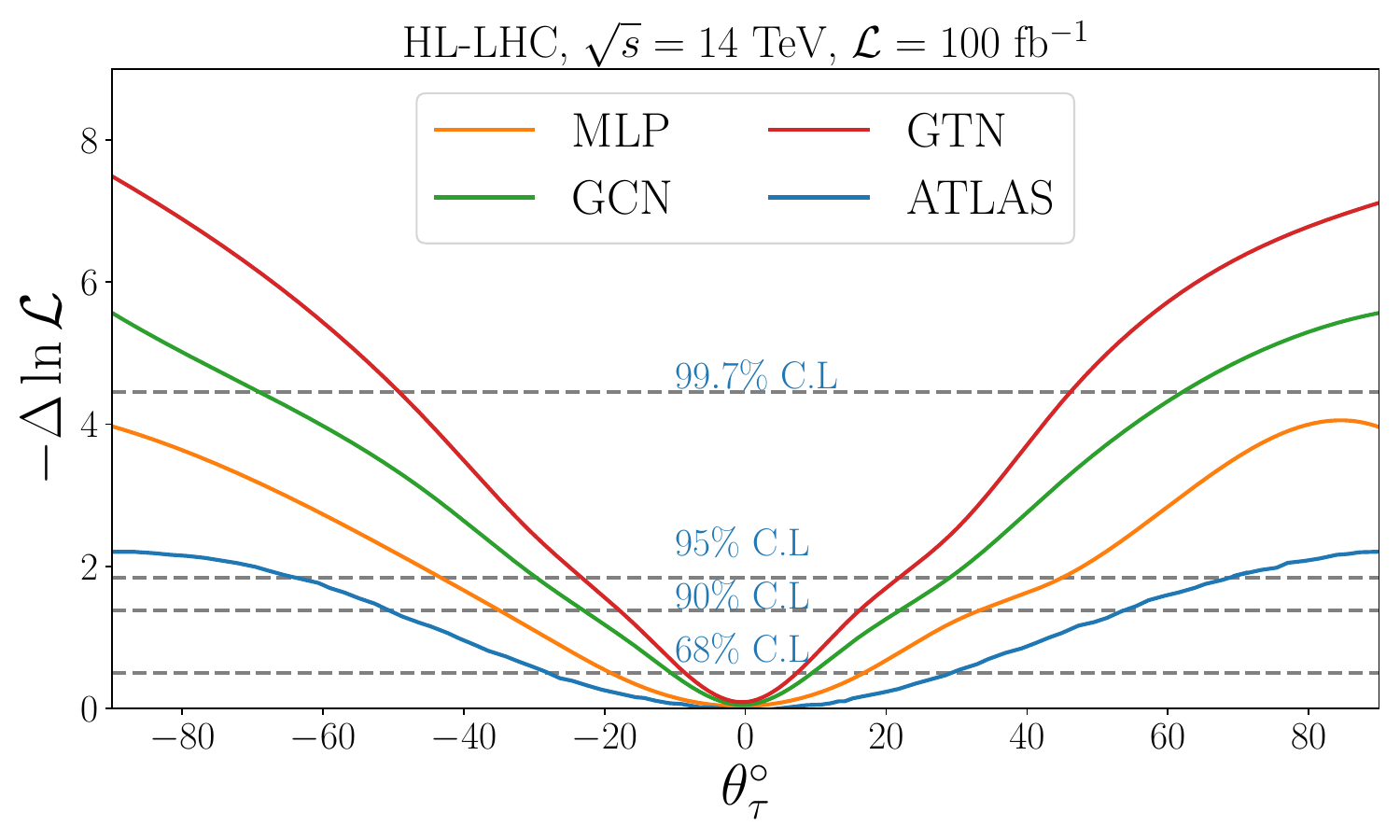}
    \caption{The binned log-likelihood result as a function of $\theta_\tau$ for the three networks, MLP (orange), GCN (green), GTN (red) and expected ATLAS results (blue). Expected ATLAS results are extracted from \cite{ATLAS:2022akr} for an analysis with  $\sqrt{s}=13$ TeV and integrated luminosity of $139 \rm fb^{-1}$. Horizontal dashed lines represent the corresponding confidence levels. }
    \label{fig:8}
\end{figure}

Distributions for the null and alternative hypothesis are weighted according to the expected number of events at the HL-LHC with $\mathcal{L}=100\rm fb^{-1}$ after imposing the cut on the network output probability to maximize the signal to background ratio. Figure \ref{fig:8} shows the binned log-likelihood for MLP (orange), GCN (green), GTN (red) and ATLAS results (blue) versus CP mixing angle. We extracted the ATLAS results from a recent analysis with $\sqrt{s} = 13$ TeV for an integrated luminosity of $139\rm fb^{-1}$ \cite{ATLAS:2022akr}. It excludes the CP mixing angle $\theta_\tau \ge |28^\circ|$ at $68\%$ C.L. A similar analysis performed by CMS found an expected value of  $\theta_\tau \ge |21^\circ|$ \cite{CMS:2022uox} at $68.3$~\% C.L with integrated luminosity of $138 \rm fb^{-1}$. 
 As seen in figure~\ref{fig:8}, MLP excludes $\theta_\tau \ge |43^\circ|$ at $95\%$ C.L. 
 GTN shows a superior performance excluding $\theta_\tau \ge |22^\circ|$ at $95\%$ C.L,  while GCN excludes $\theta_\tau \ge |31^\circ|$ at $95\%$ C.L. 
 MLP excludes the pure CP-odd states nearly $3 \sigma$ while GCN and GTN improve this further. 
The plots indicate that an improved DNN analysis can enhance both the current LHC and future HL-LHC search. 
Note that our results are obtained under significant simplification on the event reconstruction for the number of $\pi^\pm$ and $\pi^0$; therefore, they cannot be used for a direct comparison with the experimental results. 
The recent studies by ATLAS \cite{ATLAS:2022akr} and CMS \cite{CMS:2022uox} constraint the pure CP odd Higgs at $2\sigma$ level as can be seen in the figure. Our MLP results using high-level variables are comparable to the ATLAS ones, indicating the simplified analysis does not affect the core message of this paper.

\section{Conclusion}
\label{sec:6}
In this paper, we investigate the CP structure of the $h\tau^\pm\tau^\mp$ vertex at the HL-LHC with $\sqrt{s}=14$ TeV.  
We consider three different channels of hadronic $\tau$ lepton decays for the case where both of the $\tau$ lepton decays into the same final state: $\tau^\pm\to \rho^\pm (\rho^\pm\to\pi^\pm\pi^0 ) \nu_\tau$, $\tau^\pm\to a_1^\pm (a_1^\pm\to\pi^\pm\pi^0\pi^0 ) \nu_\tau$, and $\tau^\pm \to\pi^\pm\nu_\tau$.
The CP structure of the $h\tau^\pm\tau^\mp$ vertex can be determined from the angular correlation of the $\tau$ spins. 
This correlation can be reconstructed from
the angular distribution between the charged and neutral pions of the $\tau$ lepton pair decays, even though there are tau neutrinos in the final state. %

To improve the projected reach of measuring the CP mixing angle at the HL-LHC, we consider utilizing advanced deep-learning networks to enhance the signal-to-background yield. For this purpose, we employ three different networks to analyze different data structures. The MLP is used to analyze the kinematic distribution and reconstruct the CP mixing angle. 
However, due to the fully connected nature of the MLP, it fully mixes the kinematic and CP information, diluting the learned CP information and hindering the overall classification performance. To overcome this, we consider heterogeneous graphs constructed from the information stored in the final and decayed particles. With a selective connection of the nodes we fix the processing of the kinematic and CP information. For heterogeneous graph analysis, we adopt two networks, GCN and GTN. We find that GTN shows superior performance in background rejection, achieving a signal significance of $8.6\sigma$ at the HL-LHC with $\mathcal{L}=100 \, \rm{fb}^{-1}$ for a benchmark point with $\theta_\tau = 90^\circ$. GCN and MLP have lower signal significance for the same benchmark point with $7.19\sigma$ and $5.6\sigma$, respectively.

After improving the background rejection, we perform a shape analysis for the remaining events. We use a binned negative log-likelihood analysis to estimate the probability of seeing $\theta_\tau \neq 0$ at the HL-LHC. 
Keeping the limitation of simplified analysis for $\tau$ jet tagging using truth information of $\tau\rightarrow \rho$ and $a_1$ decays and $\pi^0$ reconstruction, 
our results show that the pure CP-odd state is excluded at nearly $3\sigma$ using the MLP, while GCN and GTN exclude the pure CP-odd state at above the  $3\sigma$ level showing stronger significance than current LHC measurements \cite{ATLAS:2022akr,CMS:2022uox}. Moreover, GTN shows superior performance excluding $\theta_\tau \ge |22^\circ|$ at $95\%$ C.L., while GCN excludes $\theta_\tau \ge |31^\circ|$ at $95\%$ C.L. and MLP excludes $\theta_\tau \ge |43^\circ|$ at $95\%$ C.L. 
GTN's improved performance over GCN is due to the fact that GTN applies an attention mechanism during training. The main advantage of the attention mechanism is that it assigns weights to different elements in the input graph, emphasizing the more relevant parts while downplaying the less relevant ones. Conversely, GCN treats all neighbouring vertices equally during the feature aggregation process, which can lead to suboptimal performance if certain neighbours provide more valuable information than others, as is the case encoded in the considered heterogeneous graphs.
To ensure the reproducibility of our results, we have made our codes and files publicly available on \href{https://github.com/wesmail/HiggsCP}{https://github.com/wesmail/HiggsCP}

\subsection*{Acknowledgments}
MN and AH are funded by grant number 22H05113, ``Foundation of Machine Learning Physics'', Grant in Aid for Transformative Research Areas and 22K03626, Grant-in-Aid for Scientific Research (C). WE is funded by the ErUM-WAVE project 05D2022 "ErUM-Wave: Antizipation 3-dimensionaler Wellenfelder", which is supported by the German Federal Ministry of Education and Research (BMBF). CS is supported by the Office of High Energy Physics of the U.S. Department of Energy under contract DE-AC02-05CH11231 and through the Alexander von Humboldt Foundation. CS performed part of this work at the Aspen Center for Physics, which is supported by a grant from the Simons Foundation (1161654, Troyer).
\bibliographystyle{unsrt}
\bibliography{biblo}
\end{document}